\documentclass[aps,prd,12pt,preprint,tightenlines,superscriptaddress,
amsfonts,amssymb,amsmath,byrevtex,showpacs,nofootinbib]{revtex4-1}
\usepackage{graphicx,hyperref}

\newcommand{\dR}{\mathbb R}

\newcommand{\ud}{\mathrm{d}}
\newcommand{\be}{\begin{equation}}
\newcommand{\ee}{\end{equation}}

\usepackage{amssymb,amsmath}
\usepackage{latexsym}
\usepackage{graphicx}
\usepackage{color}

\DeclareMathOperator{\sech}{sech}

\begin{document}

\title{Quantum theory of the Bianchi II model}

\author{Herv\'{e} Bergeron}
\email{herve.bergeron@u-psud.fr} \affiliation{ISMO, UMR 8214 CNRS,
Univ Paris-Sud, France}
\author{Orest Hrycyna}
\email{orest.hrycyna@fuw.edu.ol} \affiliation{National Centre for
Nuclear Research, Ho{\.z}a 69, 00-681 Warszawa, Poland}
\author{Przemys{\l}aw Ma{\l}kiewicz}
\email{pmalkiew@gmail.com} \affiliation{National Centre for
Nuclear Research, Ho{\.z}a 69, 00-681 Warszawa, Poland}
\author{W{\l}odzimierz Piechocki}
\email{piech@fuw.edu.pl} \affiliation{National Centre for Nuclear
Research, Ho{\.z}a 69, 00-681 Warszawa, Poland}

\date{\today}

\begin{abstract}

We describe the quantum evolution of the vacuum Bianchi II
universe in terms of the transition amplitude between two
asymptotic quantum Kasner-like states. For large values of the
momentum variable the classical and quantum calculations give
similar results. The difference occurs for small values of this
variable due to the Heisenberg uncertainty principle. Our results
can be used, to some extent,  as a building block of the quantum
evolution of the vacuum Bianchi IX universe.
\end{abstract}


\maketitle


\section{Introduction}

In cosmology, almost all known general relativity (GR) models of
the Universe predict the existence of cosmological singularities
with blowing up gravitational and matter field invariants. These
singularities indicate the breakdown of classical theory at
extreme physical conditions. The existence of the cosmological
singularities in solutions to GR signals incompleteness of the
classical theory. It is expected that a consistent theory of
quantum gravity should resolve the classical singularities.

The Belinskii-Khalatnikov-Lifshitz (BKL) scenario
\cite{BKL22,BKL33} is thought to be a generic solution to the
Einstein equations near spacelike singularity. It has been proved
that the isotropy of spacetime is dynamically unstable in the
evolution towards the singularity (see, e.g. \cite{BKL22,Bogo}).
Therefore, it seems that the commonly used
Friedmann-Robertson-Walker (FRW) model cannot be used to model the
very early Universe. The prototype for the BKL scenario is the
vacuum  Bianchi IX model \cite{BKL22}. The building blocks of this
model are the vacuum  Bianchi I and II models
\cite{BKL22,Reiterer:2010cz}. We expect that obtaining quantum
versions of the latter models may be helpful in the quantization
of the former one. The quantum Bianchi IX model may enable finding
the nonsingular quantum BKL theory, which could be used as a
realistic model of the very early Universe.

The quantization of the Bianchi I, II, and IX models with the aim
of resolving the classical singularity problem has already been
proposed within the loop quantization approach
\cite{Ashtekar:2009vc,Ashtekar:2009um,WilsonEwing:2010rh}. Our
investigations of the quantum Bianchi I model, based on a
modification of the latter method, can be found in
\cite{DzP,MPDz,Malkiewicz:2011sr}. The goal of the present article
is quite different. Namely, we treat the Bianchi II as a model of
a single transition between two consecutive Kasner's epochs of the
Bianchi IX dynamics only. It is not expected to be valid at the
cosmological singularity. The quantization of some anisotropic
cosmological models has been explored before (see, e.g.,
\cite{TCh,JEL}), but mostly within the Dirac approach. Our
procedure involves a reduction of the dynamical constraint at the
classical level, followed by quantization of true Hamiltonian.

In Sec. II we first present the Misner-like canonical formulation
of our homogeneous models. In this approach, the Universe is
interpreted to be a ``particle'' with its mass depending on time
and position in space \cite{cwm1,cwm2}. Next, we find dynamical
interrelations between both classical Bianchi models. The quantum
level is presented in Sec. III. We solve the Schr\"{o}dinger
equation for the Bianchi II model. The solution is interpreted in
terms of the ``scattering'' of the Kasner universe against the
potential wall of the Bianchi II universe. The asymptotic form of
the solution enables determination of the scattering amplitude. In
the last section we suggest  that our results can be used, to some
extent, as a building block of the evolution of the Bianchi IX
model.

\section{Classical dynamics}

We assume spacetime admitting a foliation $\mathcal{M}\mapsto
\Sigma\times\mathbb{R}$, where $\Sigma$ is spacelike. The line
element of the spatially homogenous, diagonal Bianchi models reads
\begin{equation}\label{eq}
\ud s^2=-N(t)^2\ud t^2+\sum_i q_i(t){\omega^i}\otimes{\omega^i}\, ,
\end{equation}
where $\omega^i$ are 1-forms on $\Sigma$ invariant with respect to
the action of a simply transitive group of motions on the leaf and
subject to
\begin{equation}\label{cartan}
\ud\omega^i=\frac{1}{2}C^i_{jk}\omega^j\wedge\omega^k\, ,
\end{equation}
where $C^i_{jk}$ are structure constants of the corresponding Lie
algebra. In the case of the Bianchi I model one has $C^i_{jk}=0$
for any $i,j,k$. The Bianchi II model is specified by the only
nonvanishing $C^1_{23}= - C^1_{32} = n_1 \ne 0$. One  can choose
$n_1>0$, and its value is usually fixed by the condition  $n_1=1$.
In what follows we keep $n_1$ as a parameter. A solution to
(\ref{cartan}) reads
\begin{equation}\label{formsol}
\omega^1=dx-n_1z dy,~~\omega^2=dy,~~\omega^3=dz ,
\end{equation}
where $x$, $y$ and $z$ are coordinates on $\Sigma$.

We recall that in canonical relativity there are the so-called
diffeomorphism constraints, which are first class and generate
canonical transformations, whose action geometrically corresponds
to spatial diffeomorphisms in space-time. Those diffeomorphisms
are viewed as coordinate transformations and, as such, are
unphysical. The gauge-fixing procedure may be applied to extract
the physical degrees of freedom \cite{HT}. The diffeomorphism
constraints, however, are absent in our model because of the
spatial homogeneity, which allowed us to fix the metric in the
form of Eq. (\ref{eq}). Nevertheless, there are still some
restricted (homogeneous) transformations of $\omega^i$'s so the
gauge is not fixed completely and we need to fix it further.

Let $\tilde{\omega}^1$, $\tilde{\omega}^2$ and $\tilde{\omega}^3$
be another solution of (\ref{cartan}). Then, because the old and
the new solutions are invariant with respect to the action of the
same homogeneity group, they must be related by a linear
transformation $L^i_{~j}$ such that
\begin{equation}
\tilde{\omega}^i=L^i_{~j}\omega^j, ~~~(L^{i'}_{~i})^{-1}C^i_{jk}L^j_{~j'}
L^k_{~k'}=C^{i'}_{j'k'}.
\end{equation}
This implies in particular that $d\tilde{\omega}^2=0=d\tilde{\omega}^3$.
Thus, $L^2_{~1}=0=L^3_{~1}$, and
\begin{equation}
\tilde{\omega}^2=L^2_{~2}\omega^2+L^2_{~3}\omega^3,~~~
\tilde{\omega}^3=L^3_{~2}\omega^2+L^3_{~3}\omega^3 ,
\end{equation}
We set $\Lambda:=L^2_{~2}L^3_{~3}-L^2_{~3}L^3_{~2}\neq 0$. From it
easily follows that
\begin{equation}
d\tilde{\omega}^1=\tilde{\omega}^2\wedge\tilde{\omega}^3=\Lambda
{\omega}^2\wedge{\omega}^3= \Lambda d\omega^1~~\Rightarrow
\tilde{\omega}^1= \Lambda
{\omega}^1+L^1_{~2}\omega^2+L^1_{~3}\omega^3 .
\end{equation}
Thus, at the level of coordinates, the possible transformation (up to a
constant shift) is the following
\begin{equation}
\tilde{x}=\Lambda x,~~\tilde{y}=L^2_{~2}y+L^2_{~3}z,~\tilde{z}
=L^3_{~2}y+L^3_{~3}z .
\end{equation}
However, when combined with the requirement that $L^i_{~j}$
preserves the form of the metric (\ref{eq}), that is
\begin{equation}
\sum_i q_i(t){\omega^i}\otimes{\omega^i}=\sum_i \tilde{q}_i(t)
{\tilde{\omega}^i}\otimes{\tilde{\omega}^i}
\end{equation}
we find that $L^2_{~3}=0=L^3_{~2}$. Now, we demand that $x$ and
$y$ be coordinates on a compact manifold, say $\mathbb{T}^2$, such
that $\int_{\mathbb{S}}dx = \int_{\mathbb{S}}dy =1$, while $z \in
\dR$. This fixes the 1-forms completely and restricts the allowed
coordinate transformations as follows:
\begin{equation}
\tilde{x}=x+x_0,~~\tilde{y}=y+y_0,~\tilde{z}=z+z_0,
\end{equation}
Thus, the variables $q_i$ are now physical.

We assume a fiducial cell which is the Cartesian product of the
whole $\mathbb{T}^2$ and any compact subset
$\sigma_z\subset\mathbb{R}$. The following holds:
\begin{equation}
\int_{\mathbb{T}^2\times\sigma_{\tilde{z}}}d\tilde{x}
\wedge d\tilde{y}\wedge d\tilde{z}=\int_{\mathbb{T}^2\times\sigma_{z}}dx\wedge dy\wedge dz
\end{equation}
so even if the variable $z$ may be chosen only up to a shift, the
volume of the patch is uniquely defined. We believe that having a
well-defined volume of the fiducial cell is essential for having
physically meaningful variables in the case of spatially
homogeneous and noncompact universes.

Due to the homogeneity of the space, the action of the {\it
vacuum} Bianchi I and II models takes the general form
\begin{equation}
\label{action} S  =  \frac{1}{\kappa}
\int_{\mathcal{V}\subset\Sigma} \, \eta \, \int N L (
q_i,\frac{\dot{q}_i}{N})  \, dt  = \frac{\Delta}{\kappa} \int N \,
dt  \, L( q_i,\frac{\dot{q}_i}{N} )\;,
\end{equation}
where (i) $N$ is the lapse function, $L$ is the Lagrangian
function of  $(q_i,\dot{q}_i)$ and $\kappa=8\pi G$, (ii) $\eta$ is
the 3-form $\eta= \omega_1 \wedge \omega_2 \wedge \omega_3$.
$\mathcal{V}$ denotes a finite patch in $\Sigma$, over which the
integration is performed. $\Delta= \int_{\mathcal{V}\subset\Sigma}
\eta$ will be  the fiducial volume in $\Sigma$.

Therefore the models only depend on the effective gravitational
constant $\tilde{\kappa}=\kappa / \Delta$. The Hamiltonians read
\cite{Bogo}
\begin{equation}\label{ham0}
H = \frac{1}{\tilde{\kappa}}
\frac{N}{\sqrt{q_{1}q_{2}q_{3}}}\bigg(-q_{1}^{2}p_{1}^{2}-q_{2}^{2}
p_{2}^{2}-q_{3}^{2}p_{3}^{2} +2  \big(q_{1}p_{1}q_{2}p_{2}+q_{1}p_{1}q_{3}p_{3}
+q_{2}p_{2}q_{3}p_{3}\big)-\frac{1}{4}n_{1}^{2}q_{1}^{2}\bigg)
\end{equation}
where $p_k$ denotes the conjugate momentum variable to $q_k$,
according to the Poisson bracket $\{q_i, p_j\}= \tilde{\kappa}\,
\delta_{ij}$. Also, $n_1 = 0$ and $n_1 \ne 0$ correspond to the
Bianchi I and Bianchi II models, respectively. Note that the
dimension of $\tilde{\kappa}$ reads $[\tilde{\kappa}] =
action^{-1} length^{-1}$, and assuming the $q_i$ variables to be
dimensionless, the dimensions of $p_i$ and $n_1$ are
$[p_i]=[n_1]=length^{-1}$.

Both systems have the dynamical constraint $H =0$. The case of
vanishing of the {\it physical} volume of the patch,
$\,V:=\Delta\cdot\sqrt{q_{1} q_{2} q_{3}} = 0 \,$, defines the
condition for the appearance of the cosmological singularity
\cite{Bogo}. Our paper does not address the singularity problem so
the expression (\ref{ham0}) is well defined. Our considerations
concern the evolution towards the singularity excluding the
singularity itself. We study possible quantum effects before
physical quantities reach their critical values, at which the big
bounce is expected to take place. The classical model, Bianchi II,
is assumed to be valid only as a model of a single transition
between two successive Kasner's universes, i.e., a patch of the
evolution towards the singularity, and is not meant to be a model
of the classical dynamics near the singularity. This is the reason
for our placing the singularity at infinity and emphasizing the
scattering picture at the quantum level. As the result, our
findings are limited to a single event of the quantum evolution:
the quantum transition between two Kasner's epochs.

It is not difficult to relate the canonical approaches developed
independently  by Bogoyavlensky, presented in his textbook
\cite{Bogo}, and by Misner \cite{cwm1,cwm2,RS}. The latter one has
been commonly used for about four decades. The former is a
comprehensive analysis of the classical dynamics of all
homogeneous models carried out within one formalism.

For further analysis we introduce Misner's like three canonical
pairs $(\beta_0, \pi_0,\beta_\pm, \pi_\pm)$ as follows
\begin{equation}\label{can1}
\beta_0 := 
\ln{(q_{1}\sqrt{q_{2} q_{3}})},~~~\beta_+
:= \ln q_1,~~~\beta_- :=
\ln{\sqrt{q_{2}/q_{3}}}
\end{equation}
\begin{equation}\label{can2}
\pi_0 := q_2 p_2 + q_3 p_3,~~~\pi_+
:= q_1 p_1 - q_2 p_2 - q_3
p_3, ~~~\pi_- := q_2 p_2 - q_3 p_3
\end{equation}

One easily verifies that $\{\beta_i, \pi_j\}= \tilde{\kappa} \,
\delta_{ij}$. The Hamiltonian (\ref{ham0}) in these new variables
has the form
\begin{equation}\label{ham2}
H = N e^{-(\beta_0 -\frac{1}{2}\beta_+) }\Big( \pi_0^2 - \pi_+^2 -
\pi_-^2 - \frac{n_1^{2}}{4} e^{2 \beta_+}\Big).
\end{equation}
We note that $\pi_0$ is a dynamical constant and that the sign of
$\pi_0$ corresponds to the direction of evolution. We will use
this fact to define the true Hamiltonian of the system(s).

\subsection{True Hamiltonian}

The canonical 2-form $\omega$ that can be ascribed to the
six-dimensional {\it kinematical} phase space of our system reads
\begin{equation}\label{b6}
\tilde{\kappa}\cdot\omega =  \ud \pi_+ \wedge \ud \beta_+ + \ud \pi_- \wedge \ud \beta_- + \ud
\pi_0 \wedge \ud \beta_0.
\end{equation}
We restrict $\pi_0>0$ and introduce the new canonical pair
\begin{equation}
\beta_0'=-\frac{\beta_0}{2\pi_0},~~\pi_0'=-\pi_0^2
\end{equation}

Reduction of the form (\ref{b6}) to constraint surface $H=0$ leads
to
\begin{equation}
\tilde{\kappa}\cdot\omega|_{H=0} =  \ud \pi_+ \wedge \ud \beta_+ +
\ud \pi_- \wedge \ud \beta_- - \ud h \wedge \ud \beta_0',
\end{equation}
where
\begin{equation}\label{ham3}
h=\pi_+^2 + \pi_-^2 + \frac{n_1^{2}}{4} e^{2 \beta_+}
\end{equation}
is the true Hamiltonian in the reduced formulation. Thus, as one
may verify that the following is satisfied:
\begin{equation}\label{eq1}
\frac{\ud\beta_\pm}{\ud \beta_0'}= \frac{\partial h}{\partial
\pi_\pm},~~~~\frac{\ud\pi_\pm}{\ud \beta_0'}= - \frac{\partial
h}{\partial \beta_\pm}.
\end{equation}
Therefore, we have the Hamiltonian system defined in the physical
phase space with $h$ being the generator of motion and $\beta_0'$
playing the role of time. Note that the direction of evolution is
set by the growth of $\beta_0'$, which  corresponds to the
contraction of universe.  Furthermore, by the virtue of Eq.
\eqref{eq1}, the classical dynamics is invariant with respect to
the choice of the fiducial cell $\mathcal{V}$ in $\Sigma$.

\subsection{Bianchi I as the asymptotic past/future of Bianchi II}

In what follows we find dynamical relation between the two Bianchi
models. This has already been done (see, e.g. \cite{Bogo}), to
some extent, but within different parametrization of phase space
and in different context. Here, we use Misner's like variables,
which are convenient in our quantization procedure. This way we
obtain the consistency between classical and quantum levels.

Equations (\ref{eq1}) read explicitly
\begin{equation}\label{eq5}
\dot{\beta}_+ =  2 \pi_+,~~~\dot{\beta}_- = 2
\pi_-,
\end{equation}
\begin{equation}\label{eq6}
\dot{\pi}_+ = -  \frac{n_1^{2}}{2} e^{2\beta_+},~~~\dot{\pi}_- =
0.
\end{equation}
The system (\ref{eq5})-(\ref{eq6}) integrates easily for the
Bianchi II case ($n_1 \neq 0$) to the form
\begin{equation}\label{eqq2}
\beta_+ = \ln{\left[\sech{\left(a_{1} \beta_0' +a_{2}\right)}\right]}+\ln
\frac{a_1}{n_1},~~~\beta_-
= a_3 \beta_0' + a_4,
\end{equation}
\begin{equation}\label{eqq3}
\pi_+ = - \frac{a_{1}}{2}\tanh\left(a_{1} \beta_0' +a_{2}\right),~~~ \pi_-
= \frac{a_3}{2},
\end{equation}
where $a_1, a_2, a_3, a_4$ are real constants, $a_1>0$. It is
clear that the dimension of $a_1$ and $a_3$ is an \emph{action}
(the inverse of the dimension of $\beta_0'$). The remaining
constants are dimensionless.  Asymptotically, as $\beta_0'
\rightarrow \pm \infty$, we obtain $\beta_+  \rightarrow - \infty$
and $\pi_+\rightarrow \mp \frac{a_{1}}{2}$. Another way of
obtaining this result  is realizing that (\ref{eq5})-(\ref{eq6})
imply
\begin{equation}\label{eqq4}
\ddot{\beta}_+ = - (n_1 e^{\beta_+})^2 <0, ~~~\beta_0' \in
\dR ,
\end{equation}
which means that the graph of $\beta_+$ is globally concave.
Consequently,  $\beta_+ \rightarrow - \infty$, as $\beta_0'
\rightarrow \pm \infty$, which is the key in showing the
asymptotic equivalence of the two Bianchi models.

In the case of the Bianchi I model ($n_1=0$), Eqs. (\ref{eq1})
read explicitly
\begin{equation}
\dot{\beta}_+ =  2 \pi_+,~~~\dot{\beta}_- = 2
\pi_-,
\end{equation}
\begin{equation}
\dot{\pi}_+ = 0,~~~\dot{\pi}_- = 0,
\end{equation}
and have the obvious solution
\begin{equation}\label{eqq5}
\beta_+ =  b_1 \beta_0' + b_2,~~~\beta_-
= b_3 \beta_0' + b_4,
\end{equation}
\begin{equation}\label{eqq6}
\pi_+ = \frac{b_1}{2},~~~ \pi_-
= \frac{b_3}{2},
\end{equation}
where $b_1, b_2, b_3, b_4$ are real constants. We note that the
Bianchi II solution (\ref{eqq2}-(\ref{eqq3}) for large $\pm
\beta_0'$ coincides with the Bianchi I solutions
(\ref{eqq5}-(\ref{eqq6}) with\begin{equation} b_1=\mp
a_1,~~b_2=\ln\frac{a_1}{n_1}\mp a_2,~~b_3=a_3,~~b_4=a_4\,.
\end{equation}
Therefore, we have explicitly shown   that asymptotically, as time
goes to $\pm \infty$, the solutions of the two Bianchi models
coincide.

\subsection{An energy-dependent wall approximation for Bianchi-II}

Using Eqs. \eqref{eqq2} and \eqref{eqq3} and introducing the
definition $\pi_+^\infty = a_1/2 >0$ for the asymptotic value of
$\pi_+$ at $\beta_0' \to -\infty$, we see that the trajectory
(qualitatively) looks like a reflection on an infinite wall. The
position of the wall can be obtained from the turning point of the
trajectory $\dot{\beta_+} =2 \pi_+= 0$. We deduce from
\eqref{eqq3} that $\beta_0'=-a_2/a_1 = -2 a_2/\pi_+^\infty$, and
the corresponding value $\beta_+^{(C)}$ of $\beta_+$, due to
\eqref{eqq2}, reads
\begin{equation}\label{wallapp0}
\beta_+^{(C)} = \ln \frac{a_1}{n_1} = \ln \frac{2 \pi_+^\infty}{n_1} \,.
\end{equation}
Thus, the main features of the trajectories can be grasped via
introducing an infinite wall approximation with the position
$\beta_+^{(C)}$ of the wall being $\pi_+^\infty$ dependent. As
will be seen later, the quantum version of the model completely
modifies the position $\beta_+^{(C)}$ of the wall for small values
of $\pi_+^\infty$.

\section{Quantum dynamics}

In what follows we apply the canonical quantization method.  The
variables of the physical phase space satisfy
\begin{equation}\label{ap1}
\{\beta_+,\pi_+\} = \tilde{\kappa} = \{\beta_-,\pi_-\},
\end{equation}
with vanishing other Poisson bracket relations (for simplicity we
use here the same notation for the Poisson bracket as in the
preceding section). To quantize the algebra (\ref{ap1}) we apply
the Schr\"{o}dinger representation
\begin{equation}\label{ap3}
\beta_\pm \rightarrow \hat{\beta}_\pm f(\beta_+, \beta_-):=
\beta_\pm f(\beta_+, \beta_-),~~~\pi_\pm \rightarrow \hat{\pi}_\pm
f(\beta_+, \beta_-) := - i \hbar_b
\frac{\partial}{\partial\beta_\pm}f(\beta_+, \beta_-),
\end{equation}
where $f \in \mathcal{H}:= L^2(\dR^2,d\beta_+ d\beta_-)$ and
$\hbar_b=\tilde{\kappa} \hbar$ corresponds to the action constant
relevant in our case. In the following calculations we set
$\hbar_b=1$ to simplify expressions.

The quantum operator $\hat{h}$ corresponding to the classical
Hamiltonian  $h$ of Eq. \eqref{ham3} reads
\begin{equation}\label{ap4}
\hat{h} =\hat{\pi}_+^2 + \hat{\pi}_-^2 + \frac{n_1^{2}}{4} e^{2
\hat{\beta}_+} .
\end{equation}

Since the Hamiltonian $\hat{h}$ is time independent, the
stationary  solution to the Schr\"{o}dinger equation
\begin{equation}\label{ap}
i\frac{\partial}{\partial \beta_0'} \Psi = \hat{h}\;\Psi
\end{equation}
can be written in the form $\Psi_E(\beta_0',\beta_-, \beta_+)=
e^{-iE\beta_0'}\psi_E(\beta_-,\beta_+)$,  with  $E
>0$, where
\begin{equation}\label{aap}
\hat{h}\psi_E = E \psi_E.
\end{equation}
Therefore, the problem reduces to the problem of solving the eigen
equation (\ref{aap}).

\subsection{The generalized eigenvectors of Hamiltonian $\hat{h}$}
We have
\begin{equation}\label{ap5}
\hat{h} = \hat{h}_- + \hat{h}_+ ,
\end{equation}
where
\begin{equation}\label{ap6}
\hat{h}_- = -\frac{\partial^2}{\partial \beta^2_-},~~~\hat{h}_+ =
-\frac{\partial^2}{\partial \beta^2_+}  + \frac{n_{1}^{2}}{4} e^{2 \beta_+}.
\end{equation}
Since we have $[\hat{h}, \hat{h}_-] = 0 = [\hat{h}, \hat{h}_+] $,
the solution to (\ref{aap}) can be presented in the form
\begin{equation}\label{ap7}
\psi_E (\beta_-,\beta_+)= \phi_{\pi_-^\infty} (\beta_-) \,\phi_{\pi_+^\infty} (\beta_+),
\end{equation}
where\footnote{The operator  $\hat{h}_+$ may have only a positive
continuous spectrum.}
\begin{equation}\label{ap71}
\hat{\pi}_-  \phi_{\pi_-^\infty} = \pi_-^\infty
\phi_{\pi_-^\infty}, \, \, \pi_-^\infty \in \mathbb{R},\;\;
\text{and}\;\; \phi_{\pi_-^\infty}(\beta_-)= e^{i \pi_-^\infty
\,\beta_-}\,,
\end{equation}
\begin{equation}\label{ap72}
\hat{h}_+ \phi_{\pi_+^\infty}  = e_+  \phi_{\pi_+^\infty}, \,\,
e_+ \ge 0, \,\, \text{and} \,\, \pi_+^\infty=\sqrt{e_+} >0\, ,
\end{equation}
\begin{equation}\label{ap73}
E=(\pi_-^\infty)^2+(\pi_+^\infty)^2\, ,
\end{equation}
and where  $\phi_{\pi_+^\infty}$, due to (\ref{ap6}), is the
solution to the eigen equation
\begin{equation}\label{ap8}
\big(-\frac{\ud^2}{\ud \beta^2_+}  + \frac{n_{1}^{2}}{4}e^{2
\beta_+}\big)\;\phi_{\pi_+^\infty} (\beta_+) = e_+
\;\phi_{\pi_+^\infty} (\beta_+)  \,.
\end{equation}

\begin{center}
\begin{figure}
\includegraphics[scale=0.5]{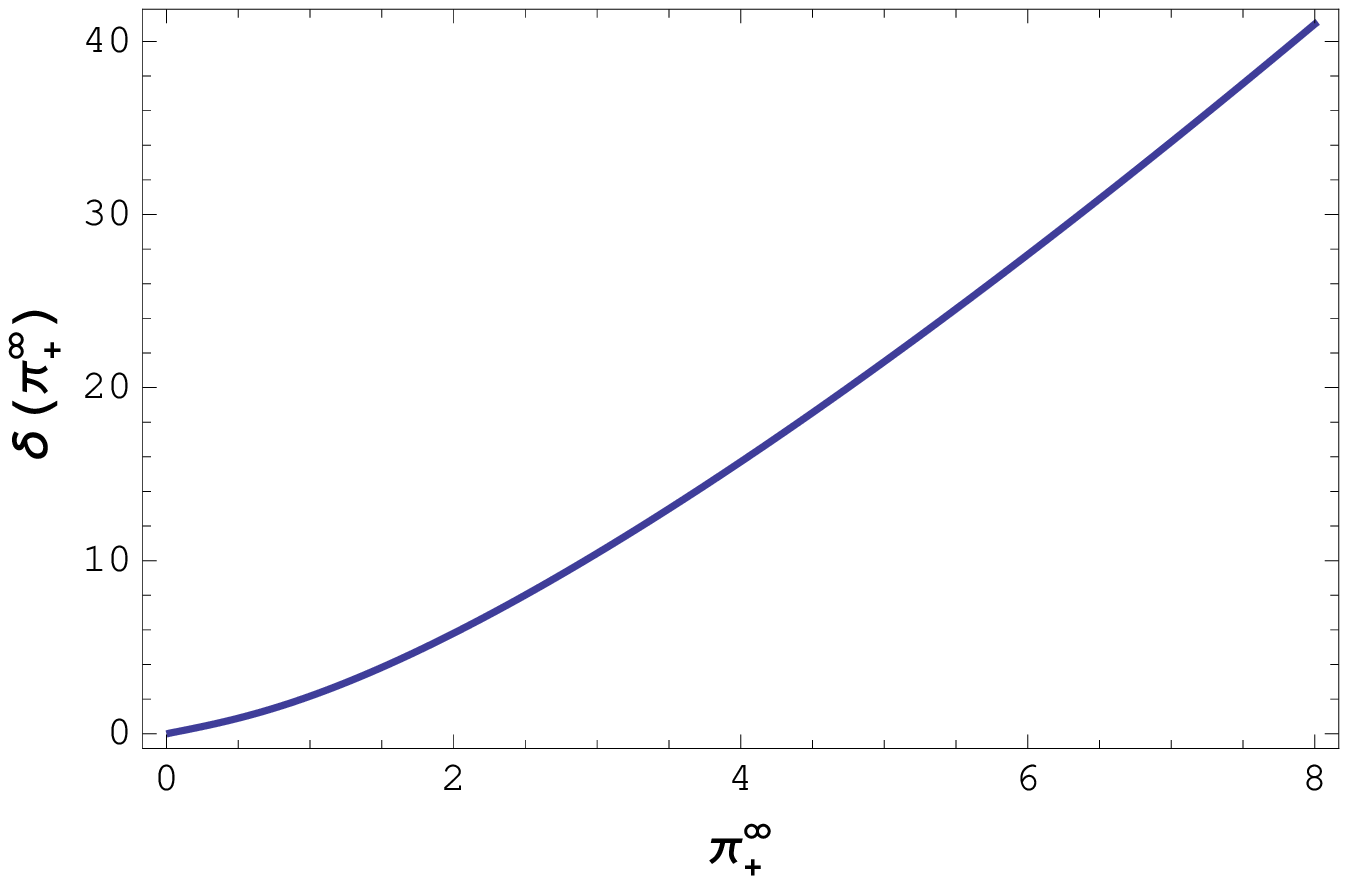}
\includegraphics[scale=0.5]{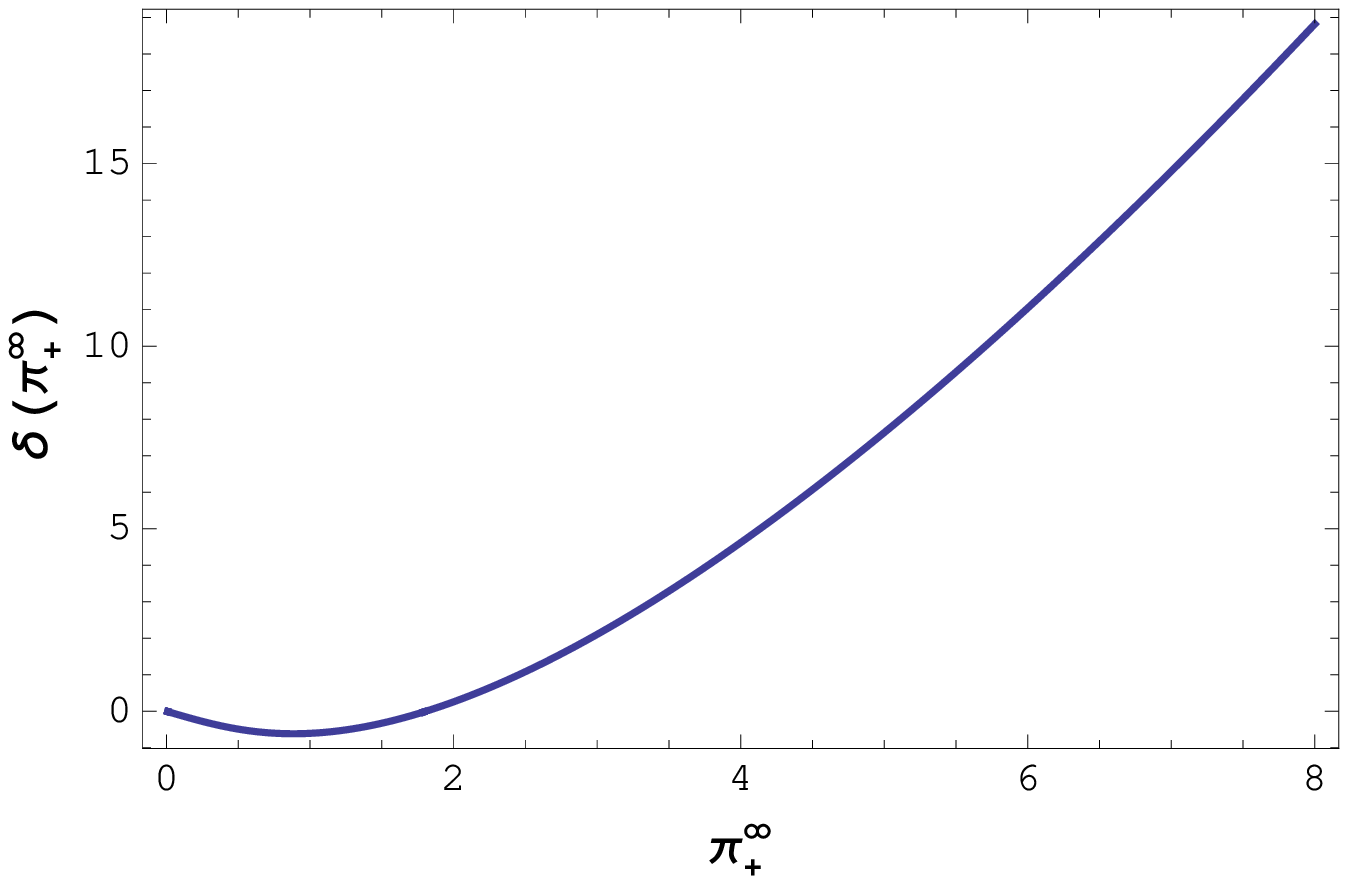}
\caption{(color online) Plot of the phase shift $\delta(\pi_+)$ of
Eq. \eqref{phasedef1} for $n_{1}=1$ (left) and for $n_{1}=4$
(right). Changing the value of $n_1$ introduces a linear additive
term.} \label{figure1}
\end{figure}
\end{center}

Equation (\ref{ap8}) has the following unique  physical solution
(no divergence for $\beta_+ \to +\infty$):
\begin{equation}
\label{eigenh3} \phi_{\pi_+^\infty} (\beta_+) = A_{\pi_+^\infty} K_{i \pi_+^\infty}
\left(\frac{n_{1}}{2} e^{\beta_+}\right) \,,
\end{equation}
where $K_\nu(x)$ are modified Bessel functions, and $A_{\pi_+^\infty}$ is
a normalization factor that can be chosen to give a suitable
behavior of $\phi_{\pi_+^\infty}$.
The complete eigenstate of $\hat{h}$ reads
\begin{equation}
\label{psiformula}
 \psi_E (\beta_-,\beta_+)= A_{\pi_+^\infty} e^{i  \pi_-^\infty \, \beta_-}
 K_{i \pi_+^\infty}
 \left(\frac{n_{1}}{2}
e^{\beta_+}\right)  \,.
\end{equation}
Using the asymptotic behavior
\begin{equation}
\beta_+ \to -\infty, \, \, K_{i \pi_+^\infty} (\beta_+) \simeq \frac{1}{2}
\left( \left( \frac{n_{1}}{4}
\right)^{i \pi_+^\infty} \Gamma(-i \pi_+^\infty) e^{i \pi_+^\infty \beta_+}
+\left( \frac{n_{1}}{4} \right)^{-i \pi_+^\infty}
 \Gamma(i \pi_+^\infty) e^{-i \pi_+^\infty \beta_+}  \right) \,,
\end{equation}
and  choosing $A_{\pi_+^\infty} :=2 (n_{1}/4)^{-i \pi_+^\infty}/ \Gamma(-i\pi_+^\infty)$,
we obtain for $\beta_+
\to -\infty$ the following asymptotic expression
\begin{equation}
\label{scatter3} \phi_{\pi_+^\infty}(\beta_+) \simeq e^{i \pi_+^\infty \beta_+} +
R(\pi_+^\infty) e^{-i \pi_+^\infty \beta_+} \quad \text{with} \quad R(\pi_+^\infty) =
 \frac{\Gamma(i \pi_+^\infty)}{\Gamma(-i \pi_+^\infty)} \left( \frac{n_{1}}{4}
 \right)^{-2 i \pi_+^\infty}\,.
\end{equation}
It corresponds to a normalized incoming wave $e^{i \pi_+^\infty
\beta_+}$ (incoming since $\pi_+^\infty >0$), and one can verify
that  $|R(\pi_+^\infty)|=1$. We interpret $R(\pi_+^\infty)$ to be
the ``reflection'' coefficient, i.e., the scattering amplitude.
Thus, the ``$S$ matrix'' transforms the asymptotic free states as
follows:
\begin{equation}
\label{smatrixh3}
 S | \pi_-^\infty,\pi_+^\infty \rangle = R(\pi_+^\infty) \,| \pi_-^\infty,
 - \pi_+^\infty \rangle \,,
 \end{equation}
 where in Dirac's notation we have $\langle \beta_-,\beta_+|\pi_-^\infty,
 \pi_+^\infty \rangle
 = e^{i \pi_-^\infty \, \beta_- } e^ {i  \pi_+^\infty \, \beta_+}$. We recover
the classical feature $\pi_+^\infty \to - \pi_+^\infty$ of the
trajectories previously studied .\\
Let us define the function $\delta(\pi_+^\infty)$ as
\begin{equation}
\label{phasedef}
R(\pi_+^\infty)= - \exp{i\delta(\pi_+^\infty)} \,.
\end{equation}
It is the ``phase shift'', where we put aside in $R$ the factor
$-1$ to represent the reflection coefficient of an infinite wall
situated at $\beta_+ =0$. In fact, due to the periodicity  $x \to
e^{i x}$, Eq. \eqref{phasedef} does not define
$\delta(\pi_+^\infty)$ uniquely. For example, if we choose
$\delta(\pi_+^\infty)= - i \ln(- R(\pi_+^\infty))$, the phase
shift $\delta(\pi_+^\infty)$ has discontinuities at
$\delta(\pi_+^\infty)=\pm \pi$. It is possible to obtain a
continuous function using the expression of the derivative
$\delta'(\pi_+^\infty)$:
\begin{equation}
\delta'(\pi_+^\infty) = -i \frac{R'(\pi_+^\infty)}{R(\pi_+^\infty)}\,,
\end{equation}
and setting
\begin{equation}
\delta(\pi_+^\infty) =\delta(0) -i \int_0^{\pi_+^\infty} \frac{R'(x)}{R(x)} \ud x \,.
\end{equation}
It is easy to see that $\lim_{\pi_+^\infty \to 0} R(\pi_+^\infty) = -1$, therefore $\delta(0)=0$, then
\begin{equation}\label{phasedef1}
\delta(\pi_+^\infty) = -i \int_0^{\pi_+^\infty} \frac{R'(x)}{R(x)} \ud x \,.
\end{equation}
Figure~\ref{figure1} presents the plot of $\delta(\pi_+^\infty)$
illustrating this case.
\subsection{Quantum energy-dependent wall approximation}

The reflection coefficient for an infinite wall situated at
$\beta_+=a$, denoted by $R_{wall}(\pi_+^\infty)$, is given by
\begin{equation}
R_{wall}(\pi_+^\infty) = -e^{2 i a \, \pi_+^\infty} \,.
\end{equation}
Therefore, we can interpret the phase shift of Eq.
\eqref{phasedef1} as being the one of a
$\pi_+^\infty$-dependent infinite wall situated at $\beta_+=\beta_+^{(Q)}$ with
\begin{equation}\label{wallapp1}
\beta_+^{(Q)} = \frac{1}{2 \pi_+^\infty} \delta(\pi_+^\infty) = -\frac{i}{2 \pi_+^\infty}
\int_0^{\pi_+^\infty} \frac{R'(x)}{R(x)} \ud x \,.
\end{equation}
Now, from $\beta_+^{(C)}$ of Eq. \eqref{wallapp0} obtained at
classical level, and $\beta_+^{(Q)}$ of Eq. \eqref{wallapp1}
obtained from quantum calculations, we obtain two different
possible approximations in terms of infinite walls. We will show
in what follows that these approximations give completely
different behavior for small values of $\pi_+^\infty$.

\subsection{The behavior of $\delta(\pi_+^\infty)$}
\begin{figure}
\begin{tabular}{cc}
\includegraphics[scale=0.5]{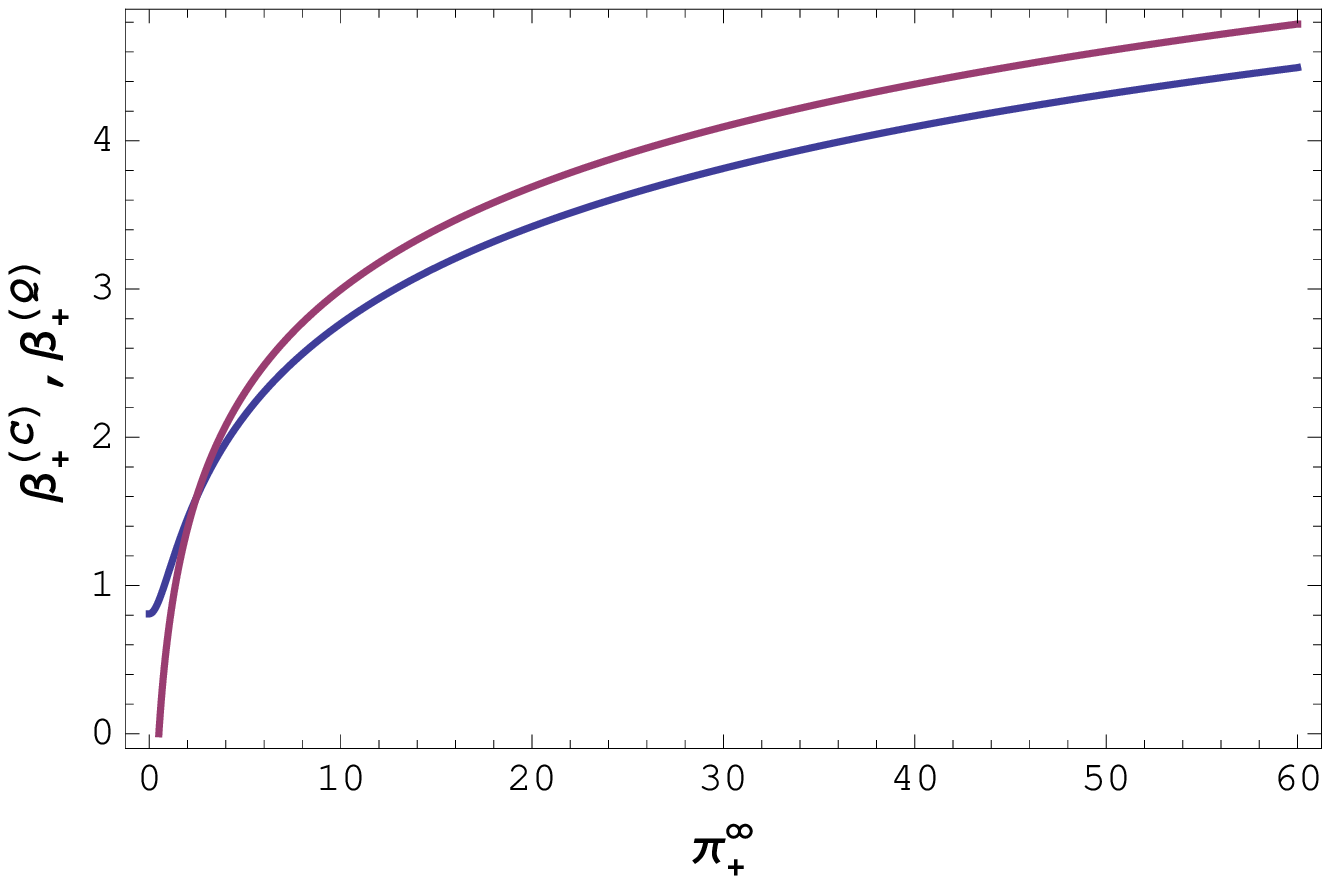}&
\includegraphics[scale=0.5]{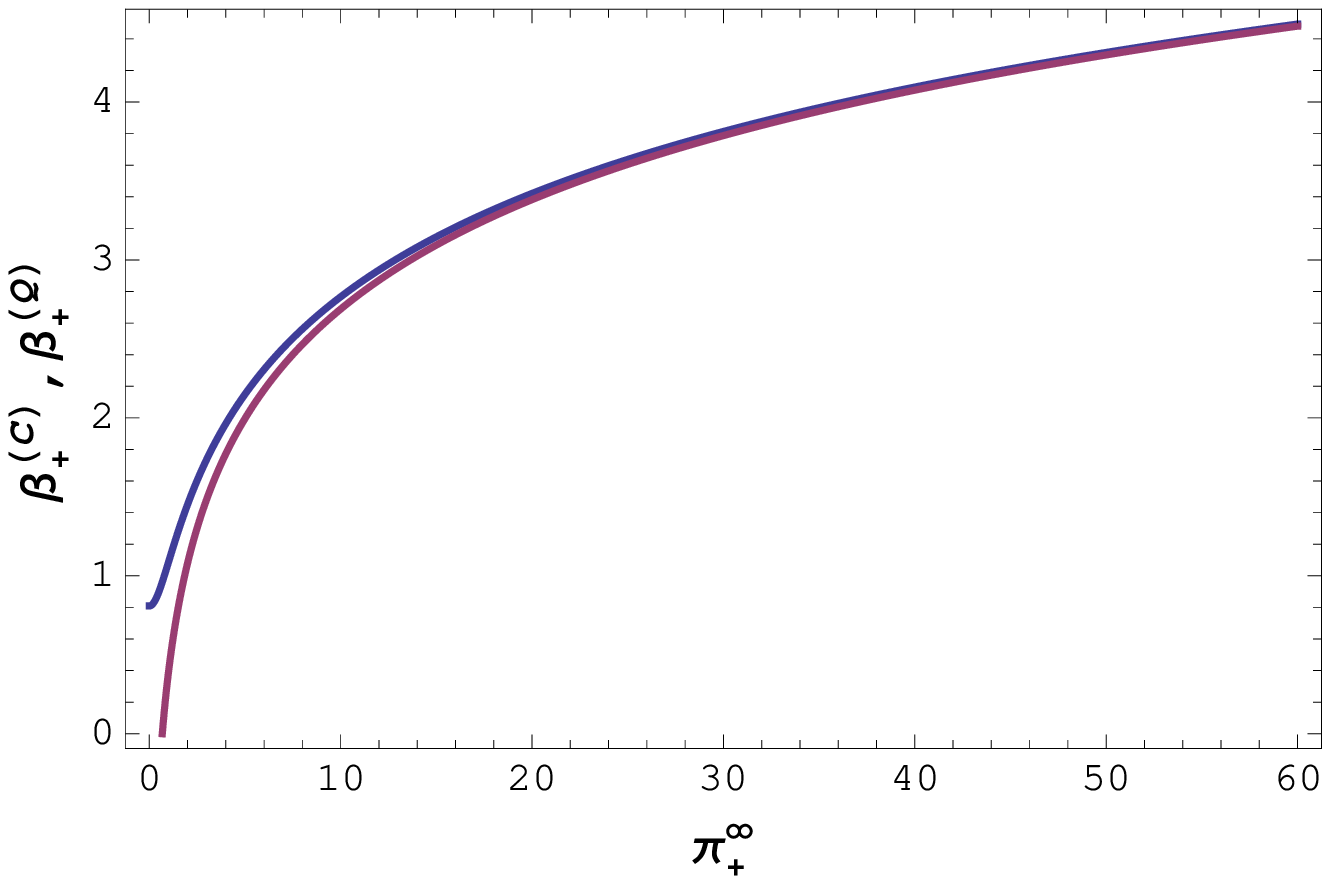}\\
\includegraphics[scale=0.5]{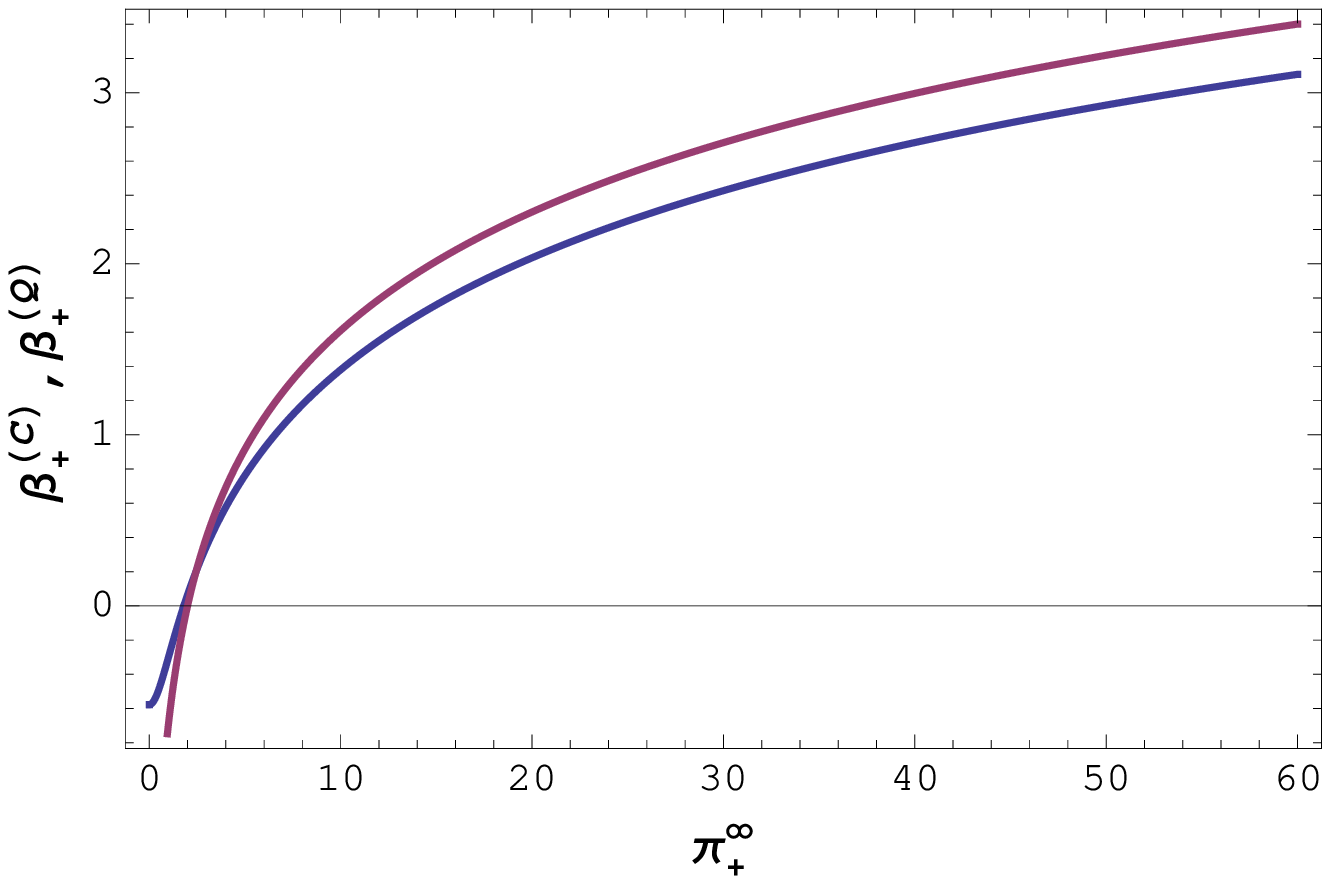}&
\includegraphics[scale=0.5]{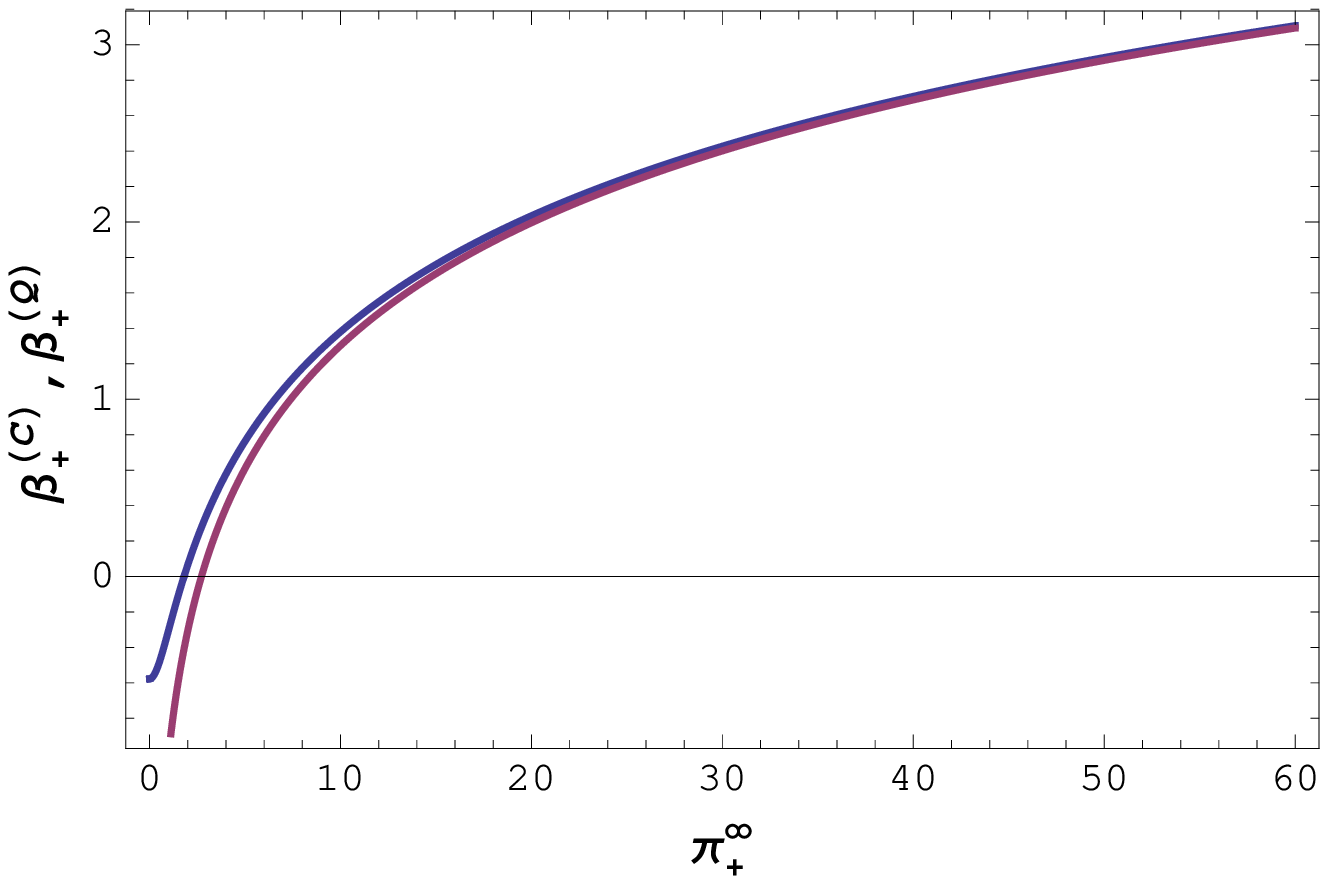}
\end{tabular}
\caption{(color online) Plots of $\beta_+^{(C)}$ (red curve) and
$\beta_+^{(Q)}$ (blue curves) for $n_{1}=1$ (top) and for
$n_{1}=4$ (bottom). Changing the value of $n_1$ introduces the
same shift on $\beta_+^{(C)}$ and $\beta_+^{(Q)}$. The left plots
are without any correction; on the right, $\beta_+^{(C)}$ has been
shifted with $\delta \beta_\infty=\ln 2/e$.} \label{figure2}
\end{figure}
\subsubsection{The case of large $\pi_+^\infty$}

For large values of $\pi_+^\infty$ we obtain
\begin{equation}
\beta_+^{(Q)} = \ln \frac{4 \pi_+^\infty}{e \, n_1} + O(1/\pi_+^\infty) \,.
\end{equation}
Therefore the dominant classical and quantum expressions are equivalent
\begin{equation}
\beta_+^{(Q)} \sim \ln \pi_+^\infty \sim \beta_+^{(C)} \,.
\end{equation}
But there remains a small difference, since $\lim_{\pi_+^\infty
\to \infty} (\beta_+^{(Q)} - \beta_+^{(C)} ) =\delta \beta_\infty=
\ln 2/e$. This is shown in Fig. \ref{figure2}. This shift $\delta
\beta_\infty$ is a methodological bias that does not contain any
physical meaning. Actually we chose some ``reasonable''
definitions of the locations $\beta_+^{(C)}$ and $\beta_+^{(Q)}$
of the classical and quantum walls.  Therefore, it is not
surprising that we find finally some small numerical difference.
In the remainder we introduce the modified classical location
$\tilde{\beta}_+^{(C)} = \beta_+^{(C)} + \delta \beta_\infty$,
corresponding to the nonvanishing asymptotic part of
$\beta_+^{(Q)}$. We conclude that for large values of
$\pi_+^\infty$, classical and quantum calculations give similar
results.

\subsubsection{The case of small $\pi_+^\infty$}

\begin{center}
\begin{figure}
\includegraphics[scale=0.5]{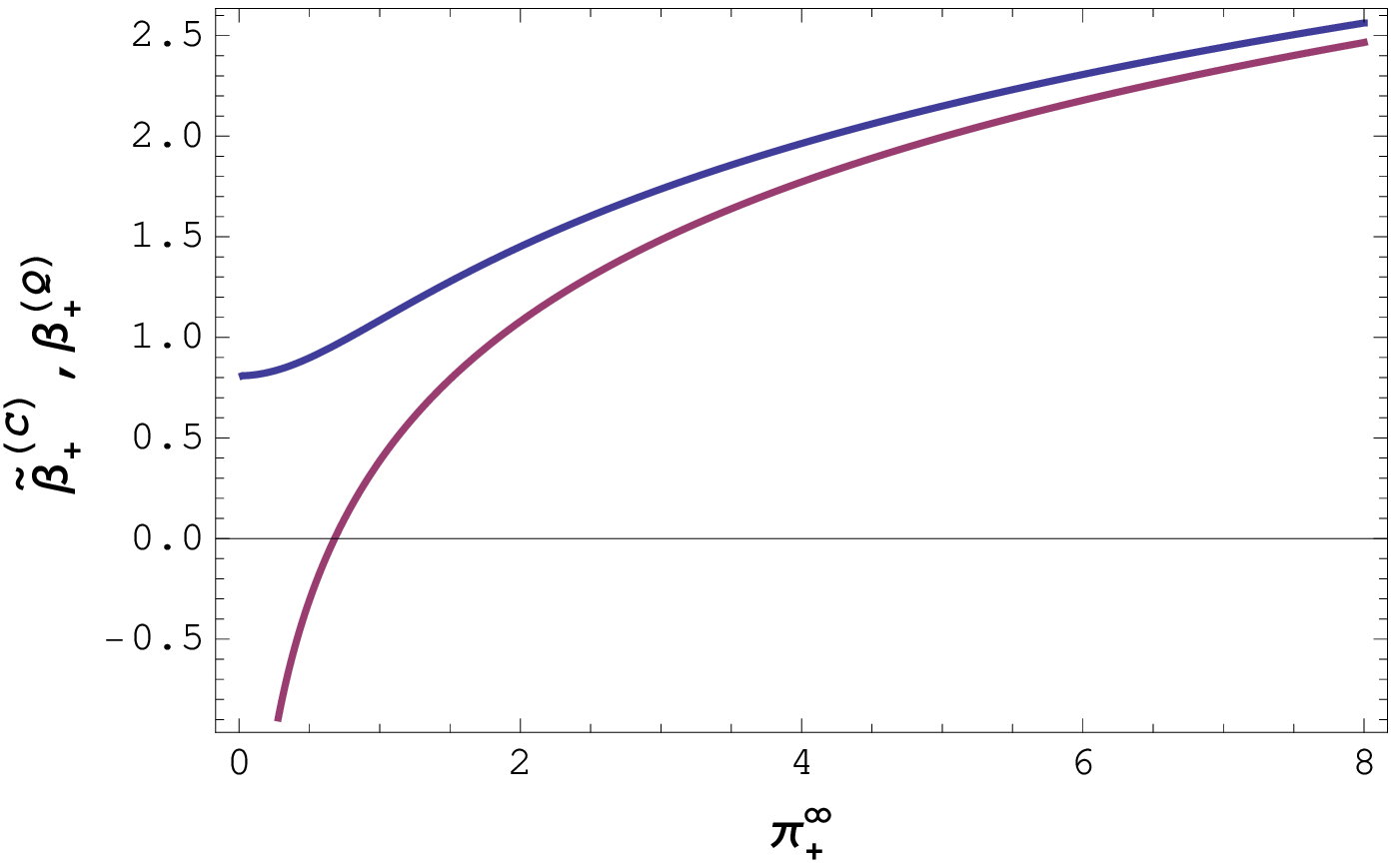}
\includegraphics[scale=0.5]{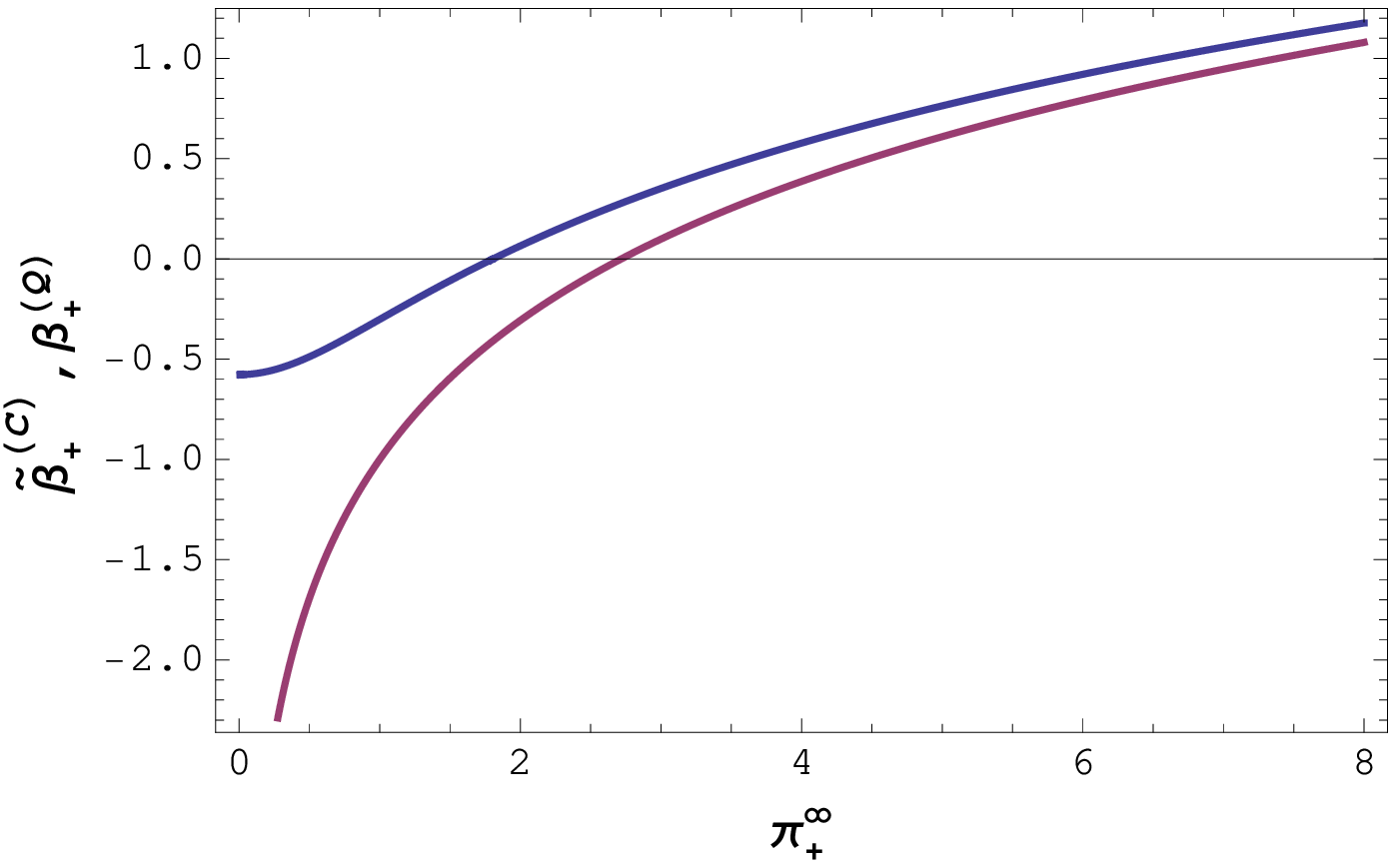}
\caption{(color online) Plots of $\tilde{\beta}_+^{(C)}$ (red
curve) and $\beta_+^{(Q)}$ (blue curve) for $n_{1}=1$ (left) and
for $n_{1}=4$ (right) and small values of $\pi_+^\infty$.}
\label{figure3}
\end{figure}
\end{center}

Making the power series  of $\delta(\pi_+^\infty)$ defined in Eq.
\eqref{phasedef1}, near $\pi_+^\infty=0$, we obtain
\begin{equation}
\delta(\pi_+^\infty) = -2 (\gamma + \ln(n_{1}/4)) \pi_+^\infty +  O((\pi_+^\infty)^3) \,,
\end{equation}
where $\gamma \simeq 0.5772$ is the Euler-Mascheroni constant.
Therefore we obtain from Eq. \eqref{wallapp1}
\begin{equation}
\beta_+^{(Q)} = - (\gamma + \ln(n_{1}/4)) + O((\pi_+^\infty)^2) \,,
\end{equation}
while from Eq. \eqref{wallapp0}, $\tilde{\beta}_+^{(C)} = \ln(4
\pi_+^\infty/e \, n_1)$
and then $\tilde{\beta}_+^{(C)} \to -\infty$ when $\pi_+^\infty \to 0$
(see Fig. \ref{figure3}).\\
Furthermore, we find $\beta_+^{(Q)} \simeq - (\gamma +
\ln(n_{1}/4))$ up to the second order. Therefore, an infinite wall
located at a fixed position $\beta_+^0=- (\gamma + \ln(n_{1}/4))$
is a very good approximation (at the quantum level), for small
values of $\pi_+^\infty$. This is a pure quantum result that does
not possess any classical counterpart. But if we transfer this
picture in the classical domain (to obtain a semiclassical
description), this means that for small values of $\pi_+^\infty$,
the accessible $\beta_+$ domain is defined by $\beta_+ \le
\beta_+^0$, which corresponds in the old variable
$q_1=e^{\beta_+}$ to the constraint
\begin{equation}
q_1 \le \frac{4 \hbar\,  \tilde{\kappa}\,  e^{-\gamma}}{n_1} \,.
\end{equation}
To restore the $\hbar_b$ dependence,  we only have to change
$\pi_+^\infty \to \pi_+^\infty / \hbar_b$ and $n_1 \to
n_1/\hbar_b$ in the expressions of $R$, $\delta$, and
$\beta_+^{(Q)}$. Therefore, the classical limit $\hbar_b \to 0$
corresponds to the previous analysis $\pi_+^\infty \to \infty$,
and we recover the asymptotic equality between classical and
quantum prescriptions. The special value $\beta_+^0$ reads
$\beta_+^0 = - (\gamma + \ln(n_{1}/4 \hbar_b))$; then, when
$\hbar_b \to 0$, $\beta_+^0 \to -\infty$ and we recover the
classical result $\lim_{\pi_+^\infty \to 0} \tilde{\beta}_+^{(C)}
= -\infty$.

\section{Summary and outlook}

We have shown explicitly the asymptotic equivalence of  classical
Bianchi I and II models: as time goes to $\pm \infty$, the
solutions to the dynamics of the two models coincide. This
circumstance was used to consider the quantum dynamics of the
Bianchi II model in terms of scattering process.  However, the
interpretation of the Hamiltonian (\ref{ap4}) as that of a
particle with mass dependent on its position in
space\footnote{That might be applied within Misner's approach
\cite{cwm1,cwm2,RS}.} is quite formal. The cosmological
interpretation is that  the Kasner universe approaching the
singularity undergoes a rapid external push to its scale factors
due to the  intrinsic curvature of the Bianchi II model.

The results presented, for small and large values of
$\pi_+^\infty$, exhibit the main common features and differences
between classical and quantum formulations. As expected,
differences become important when the nonlocality due to quantum
mechanics cannot be neglected, i.e. when the uncertainty $\Delta
\beta_+ \sim 1/\pi_+^\infty$ is very large. The Bianchi II model
is important because it bridges the two Kasner universes.  We
found that the quantization of the  Bianchi II dynamics leads to a
limited departure from the classical picture.

Roughly speaking, the classical evolution of the Bianchi IX model
\cite{BKL22} towards the cosmological singularity can be
considered to be a sequence of transitions from one Kasner epoch
to another one via the vacuum Bianchi II type evolution. This
sequence can be divided into eras which differ from one another by
oscillations of distances along different pairs of generalized
Kasner's axes \cite{BKL22,Reiterer:2010cz}. We are aware that the
classical picture needs not extend to the quantum level. We expect
that  the quantum dynamics of the Bianchi II model presented in
this paper may be used, to some extent,  as a building block for
quantum evolution of the Bianchi IX model to be examined in the
near future. Such procedure would consist in finding the suitable
way of sewing together two consecutive quantum Bianchi II models.

One the important question remainsto be answered: How can we deal
with the classical singularity of the Bianchi II model at the
quantum level? As far as we are aware, this problem has not been
addressed satisfactorily yet. Some  kernels of the dynamical
constraints of the Bianchi class A models have been found, but the
Hilbert spaces based on them have not been constructed
\cite{TCh,JEL}. We plan to address this issue elsewhere.

\acknowledgments

The research of O.H. was funded by the National Science Centre
(NSC) through the postdoctoral internship award (Decision
No.~DEC-2012/04/S/ST9/00020). The research of P.M. was supported
by the NSC Grant No. DEC-2013/09/D/ST2/03714.


\begin{thebibliography}{99}

\bibitem{BKL22} V. A. Belinskii, I. M. Khalatnikov, and E. M. Lifshitz,
Adv. Phys. {\bf 19},  525(1970).

\bibitem{BKL33} V. A. Belinskii, I. M. Khalatnikov, and E. M. Lifshitz,  Adv. Phys.  {\bf 31},  639 (1982).

\bibitem{Reiterer:2010cz}
  M.~Reiterer and E.~Trubowitz,
  arXiv:1005.4908; R.~Galimova, arXiv:1403.2767.

\bibitem{cwm1} C. W. Misner,
Phys.\  Rev.\  {\bf 186},  1319 (1969).

\bibitem{cwm2} C. W. Misner,  in {\it Magic Without
Magic: John Archibald Wheeler}, edited by J. R. Klauder (W.H.
Freeman and Company, San Francisco, 1972), p. 441.

\bibitem{Bogo} O. I. Bogoyavlensky, {\it Methods in the Qualitative Theory
of Dynamical Systems in Astrophysics and Gas Dynamics}
(Springer-Verlag, Berlin, 1985).

\bibitem{Ashtekar:2009vc}
  A.~Ashtekar and E.~Wilson-Ewing,
  Phys.\ Rev.\ D {\bf 79}, 083535 (2009).

\bibitem{Ashtekar:2009um}
  A.~Ashtekar and E.~Wilson-Ewing,
  Phys.\ Rev.\ D {\bf 80}, 123532 (2009).

\bibitem{WilsonEwing:2010rh}
  E.~Wilson-Ewing,
  Phys.\ Rev.\ D {\bf 82}, 043508 (2010).

\bibitem{RS} M. P. Ryan, Jr., and L. C. Sheply, {\it Homogeneous Relativistic
Cosmologies} (Princeton University Press, Princeton, NJ, 1975).

\bibitem{DzP}
  P.~Dzierzak and W.~Piechocki,
  Phys.\ Rev.\ D {\bf 80}, 124033 (2009).

\bibitem{MPDz}
  P.~Malkiewicz, W.~Piechocki and P.~Dzierzak,
  Classical  Quantum  Gravity  {\bf 28},  085020 (2011).

\bibitem{Malkiewicz:2011sr}
  P.~Malkiewicz,
  Classical Quantum Gravity  {\bf 29}, 075008 (2012).

\bibitem{TCh} T. Christodoulakis, G. Kofinas, E. Korfiatis, and A. Paschos,
 Phys. Lett. B {\bf 390}, 55 ( 1997).

\bibitem{JEL} J. E. Lidsey,
Phys. Lett. B {\bf 352},  207 ( 1995).

\bibitem{HT} M. Henneaux and C. Teitelboim, {\it Quantization of Gauge
Systems} (Princeton University Press, Princeton, NJ, 1992).

\end{thebibliography}
\end{document}